# The Tribology of Sliding Elastic Media

Dinko Cule and Terence Hwa
*Physics Department, University of California at San Diego, La Jolla, CA 92093-0319*

(March 12, 1996)

The tribology of a sliding elastic continuum in contact with a disordered substrate is investigated analytically and numerically via a bead-spring model. The deterministic dynamics of this system exhibits a depinning transition at a finite driving force, with complex spatial-temporal dynamics including stick-slip events of all sizes. These behaviors can be understood completely by mapping the system to the well-known problem of a directed-path in *higher-dimensional* random media.

Pinning phenomena occur widely in nature and are important to many technological applications [1]. For instance, the pinning of flux lines suppresses dissipation in a Type-II superconductor, and the pinning of contact lines controls the spreading of a liquid on a solid. These are examples of driven elastic systems embedded in higher-dimensional random media. There have been much research activities in this area in recent years: The macroscopic properties of these systems, herefoth referred to as type (i) systems, have been elucidated through a combination of the functional renormalization-group (FRG) analysis [2,3] and numerous experimental [4] and numerical [5–7] studies. In the extreme nonequilibrium limit where thermal fluctuations can be neglected and the dynamics is deterministic, one finds that a driving force exceeding a critical value is necessary to depin the system. A dynamic phase transition occurs at the depinning threshold, where the system exhibits complex stick-slip motion with avalanches of all sizes. The complex dynamics occurring in type (i) systems results from the intricate balance of elasticity and random pinning forces near the onset of motion. Several universality classes classified by symmetries have been unveiled [8,9].

Another type of pinning phenomena which have attracted much interest is the friction which occurs between two sliding bodies interacting via a contact surface. This is exemplified by the Burridge-Knopoff model of earthquake fault dynamics [10–12]. Other examples include boundary layer lubrication [13], and recent experiments on forced motion of elastic continuum (rubber and gel) over sticky substrates [14]. This type of systems, referred to here as type (ii), also exhibit complex spatio-temporal dynamics, in the limit of very slow macroscopic motion. However, the dynamics of these systems are not as well understood. Recent numerical studies [11,12] of type (ii) systems rely on certain (chaotic) instabilities of the deterministic dynamics to generate complexities. A systematic description of spatio-temporal chaos is of course notoriously difficult.

Since the interfaces of sliding bodies are almost always rough and irregular, and quenched randomness is known to generate complex dynamics (in type (i) systems), it is worthwhile to investigate the possible role of randomness in the dynamics of type (ii) systems. In this letter, we report analytical and numerical studies of a deterministically driven elastic system in contact with a *disordered* substrate. Our work is inspired by recent experimental studies of friction of elastic continuum (stretched latex membranes and gel) on sticky substrates [14]. We model the *inhomogeneous* elastic medium by an array of beads connected by *random* springs. For simplicity, we focus on the properties of the 1D version of such a random bead-spring array (i.e., a random chain), in contact with a disordered substrate (see Fig. 1). The effect of the 2-dimensional nature of the actual contact surface and the transverse dimension of the 3D elastic medium can be addressed by straightforwardly generalizing the results of the current study [15]. In our model, we assume that there is a large normal force confining the chain to the substrate, thus suppressing the transverse degrees of freedom [16]. We show below that our model of a random chain on a random substrate has a simple continuum limit, which we conjecture to be that of a directed path (DP) in 2-*dimensional* random media. The DP is the prototypical example of a type (i) systems for which much are known [1,17]. We demonstrate numerically that both the low temperature static properties of the random chain and its zero-temperature driven dynamics near the depinning threshold are indistinguishable from that of a randomly pinned DP.

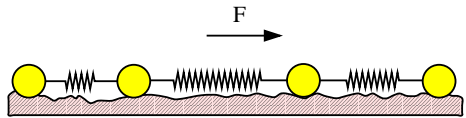

FIG. 1. The model of a random chain being pushed across a disordered substrate.

We study a one dimensional chain of beads coupled to their nearest neighbors by harmonic springs of random equilibrium lengths (see Fig. 1). The chain interacts with a (independently) random substrate described by the potential $V(x)$, with $\overline{V} = 0$, $\overline{V(x)V(x')} = \Delta_V \delta(x - x')$.

Here, the overbar denotes disorder average. The (potential) energy of the chain with $N$ beads is given by

$$E\left[\{r_n\}\right] = \sum_{n=1}^{N} \left[\frac{\kappa}{2}\left(r_{n+1} - r_n - a_n\right)^2 + V\left[r_n\right]\right], \quad (1)$$

where $r_n$ denotes the position of the $n^{th}$ bead. The elasticity constant $\kappa$ is taken to be equal for all springs. The spring lengths $a_n$ are randomly chosen from a narrow and bounded distribution with mean $a$, so that $a_n = a\left(1 + \eta_n\right)$, $\eta_n \ll 1$, with $\overline{\eta_n} = 0$, $\overline{\eta_n \eta_{n'}} = \Delta_a \delta_{n,n'}$.

We consider deterministic and purely dissipative dynamics of the system driven by a uniform external force $F$ along the chain direction. In the strongly damped limit, the equations of motion is simply given by gradient descent, i.e.,

$$\mu_0^{-1} \partial_t r_n(t) = -\frac{\delta E}{\delta r_n} + F, \quad (2)$$

where $\mu_0$ is a kinetic coefficient.

As we will soon show, the chain is completely pinned by the random potential at small driving forces. As the force increases above some threshold value, $F_c$, the chain starts to move with a nonzero average velocity, $\overline{v} \sim (F - F_c)^\beta$, for $F \gtrsim F_c$ in the steady state. As is usual in critical phenomena, the behavior of the system near the (depinning) transition is characterized by a diverging correlation length, $\xi \gg a$. It then becomes useful to *coarse grain* the discrete system (2). This will allow us to systematically filter out the microscopic details and focus on a simpler continuum equation which preserves all features relevant to the critical phenomena. The coarse-grained density and currents are defined as

$$\rho(x,t) = a^{-1} \int_{x-a/2}^{x+a/2} dx' \sum_n \delta\left(x' - r_n(t)\right), \quad (3)$$

$$j(x,t) = a^{-1} \int_{x-a/2}^{x+a/2} dx' \sum_n \partial_t r_n \delta\left(x' - r_n(t)\right). \quad (4)$$

It is convenient to introduce a displacement-like field $u(x,t)$, defined as $\rho(x,t) = \rho_0 \left(1 - \partial_x u\right)$, where $\rho_0 = 1/a$ is the average bead density. Then the continuity equation $\partial_t \rho(x,t) + \partial_x j(x,t) = 0$ becomes $\partial_t u = j(x,t)/\rho_0$. Using the equation of motion (2) in Eq. (4), then applying the coarse graining procedure and keeping the most relevant terms [15], we obtain:

$$\mu_0^{-1} \partial_t u = K\, \partial_x^2 u - \mu_0^{-1} v \partial_x u + f[u,x] + F, \quad (5)$$

where $K = \kappa a^2$ is the chain stiffness, $v(x,t) = \mu_0 (K\, \partial_x^2 u + f[u,x] + F)$ is the instantaneous velocity field as defined from $j(x,t) \equiv v(x,t)\rho(x,t)$, and

$$f = -\partial_x V + 2\rho_0 \tilde{V}_{\rho_0}(x) \sin\left[2\pi\rho_0\left(x - u - \gamma(x-u)\right)\right] \quad (6)$$

results from the coarse-graining of the pinning force $-\partial V/\partial r_n$. In (6), the sine term reflects the *discrete*

nature of the bead-spring system [15], $\tilde{V}_{\rho_0}$ denotes the $\rho_0^{th}$-Fourier component of $V$, and $\gamma(s) = \int^s ds' \eta(s')$ originates from the random spring lengths.

For a homogeneous chain (for which $\eta = 0$), Eqs. (5) and (6) become similar to the well known random-amplitude sine-Gordon equation studied in the context of charge-density waves (CDW) [2], vortex array [18], and crystal growth [19]. The discrete translational symmetry $u \to u +$ integer reflects in this case the invariance of the underlying discrete system upon a shift in the index $n$ labeling the beads. Compared to the usual CDW equation of motion, Eqs. (5) and (6) contain in addition the $u$-independent driving force $\partial_x V$ and the convective term linear in $\partial_x u$. These terms have recently been introduced on phenomenological grounds [20,21], and their effects have been investigated in the limit of strong drive.

The introduction of random springs leads to a factor $\gamma(x-u)$ which modulates the *period* of the sine term and hence breaking the discrete translational symmetry. This results from the non-uniformity of springs which destroys the relabeling invariance of the system [22]. The form of the probability distribution for $f[u,x]$ is obviously complex. However, its leading moments are straightforward to compute since the random variables $V$ and $\eta$ are uncorrelated. One finds $\overline{f[u,x]} = 0$, and

$$\overline{f[u,x] f[0,0]} \sim -\Delta_V \partial_x^2 \delta(x) \exp\left[-2\pi^2 \Delta_a \rho_0 |u|\right]. \quad (7)$$

Thus correlations in the pinning force $f[u,x]$ are short-ranged in *both* $x$ and $u$. [There is also an $u$-independent term $(\partial_x V)$ which is however irrelevant.] The dynamics of the random chain (5) (with the $\partial_x u$ term neglected) now look very much like the zero-temperature driven dynamics of a directed path (DP), described by the transverse coordinates $u(x)$, in an effective *two-dimensional* random potential $W[u,x] = -\int^u du' f[u',x]$ [3]. The convective term in (5) can be a relevant perturbation [9]. However, close to the depinning transition, it is irrelevant since $v \to 0$. There, Eq. (5) taken on the simple form $\mu_0^{-1} \partial_t u = -\frac{\delta \mathcal{H}}{\delta u}$, with $\mathcal{H} = \int dx \left[\frac{K}{2}(\partial_x u)^2 + W[u,x] - Fu\right]$, $\overline{W} = 0$ and

$$\overline{W[u,x] W[u',x']} \sim \Delta_V \Delta_a \delta(x-x') \delta(u-u'). \quad (8)$$

This result suggests that the critical behavior of the random chain may belong to the DP universality class, if the complicated correlations which exist in higher moments of the effective random potential are irrelevant [23].

To check our expectations, we performed numerical study of the bead-spring system. We first investigate the static behavior of the random chain in equilibrium using the transfer-matrix technique [15,17]. The system is simplified by discretizing the position of the beads. In Fig. 2 we present numerical data obtained from simulations of the system with $N = 4096$ beads. Disorder averaging was performed on 2000 different realizations of $V(x)$ and $a_n$. The potential $V$ was chosen from a Gaussian distribution with zero mean and unit variance. The spring lengths $a_n$ were uniformly distributed in the interval $[5, 15]$, and spring constants were kept equal at $\kappa = 0.2$. From the preceding analysis, we expect the displacement field $u_n \equiv r_n - \sum_m^n a_m$ to behave as the transverse coordinate of a directed path in a 2-dimensional random potential, with the total energy of the path $E$ given by Eq. (1). Our numerical results for the fluctuations in the net displacement $u_n$ and the total energy $E_n^*$ of the *optimal* path clearly indicate scaling (see Fig. 2), with $\overline{u_n^2} \sim n^{4/3}$ and $\Delta E_n^* \sim n^{1/3}$ at zero temperature. Similar results are obtained at finite temperatures [15]. These scaling laws are characteristic of the 2-dimensional DP universality class [17].

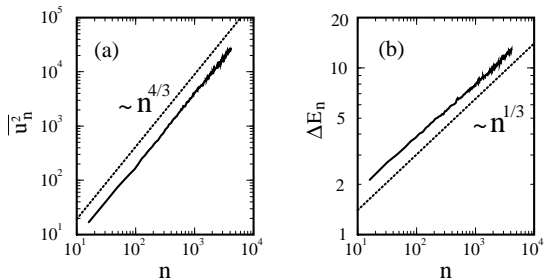

FIG. 2. Fluctuations in (a) displacement, and (b) energy of the optimal path (measured with respect to a reference path which starts and ends at $u = 0$ [15]).

We next study the zero-temperature response of the random chain to an external driving force by numerically integrating the equation of motion (2). (Here, $r_n$'s are kept as continuous variables.) The random potential $V$ used is consisted of a series of bumps and valleys similar to those used in Ref. [24]. Each has a range $b$ with an amplitude uniformly distributed in the interval $[-V_0, V_0]$. In the dynamics, we explicitly put in an additional restriction which eliminates the crossing of beads [15]. This speeds up the dynamics without affecting the asymptotic scaling properties. We consider open chains with free boundary conditions. The simulations were run for different chain lengths with parameters $\kappa = 0.1$, $b = 1$, and $V_0 = 1$. Various values of $\mu_0$ (time step size), ranging from 0.5 to 0.01, were used; we always checked that twice smaller $\mu_0$ does not change results significantly.

With a strong enough applied force $F$, the system moves and reaches a steady state after some time. The overall motion is defined by the steady state center of mass velocity $\overline{v}(F) = \frac{1}{N}\sum_{n=1}^{N}\overline{\langle v_n(t) \rangle}$, where $\langle \cdots \rangle$ denotes here a time average for a given realization of the disorder. Averaging over long time is particularly necessary in the vicinity of the depinning threshold, $F \to F_c^+$, since the motion of the chain there becomes very "jerky", exhibiting "avalanches" in which some sections of the chain move forward with much higher velocity than oth-ers. Simulations were done for system size $N = 1024$, first for constant spring lengths $a_n = 10$, and then for random springs, with $a_n$ randomly chosen from the interval $[5, 15]$. Depending on the value of $F$, time averages were taken over the intervals of $10^5$ to $10^7$ steps. Simulations were repeated several times for different $V$'s and $a_n$'s; the average results for the velocity-force characteristics are shown in Fig. 3. The best power law fit to $v \sim (F - F_c)^\beta$ gives $F_c \simeq 0.275$, $\beta \simeq 0.41$ for the uniform springs, and $F_c \simeq 0.289$, $\beta \simeq 0.25$ for random springs.

The data presented in Fig. 3 clearly shows the difference between the case of uniform and random springs, as anticipated from the above analysis. In the case of uniform springs, our estimate for exponent $\beta$ is comparable with the result $\beta = 0.45 \pm 0.05$ obtained by Myers and Sethna [25] in their simulations of an automaton model developed to study the depinning transition in CDW systems. Our result is also consistent with the FRG result of Narayan and Fisher [2] where $(4 - \epsilon)$-expansion yields $\beta = 1/2$. On the other hand, simulations of the random chain model give an exponent $\beta$ which is close to the ones obtained in recent numerical studies of the driven DP in 2-dimensional random media, where $\beta = 0.24 \pm 0.1$ [6] and $\beta = 0.25 \pm 0.03$ [7] were reported.

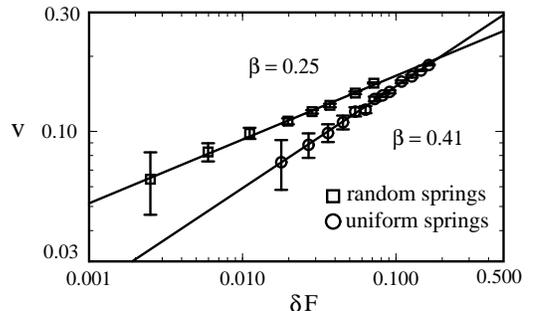

FIG. 3. Velocity-force characteristics $v \sim (\delta F)^\beta$, where $\delta F = (F - F_c)/F_c$ is the reduced driving force.

In addition to the velocity-force characteristics, we analyzed the scaling of spatial correlations. Equal-time correlations between two beads separated by $\ell$ springs are given by:

$$C^2(\ell, N) = N^{-1} \sum_n \overline{(r_{n+\ell} - r_n - a\ell)^2}. \qquad (9)$$

We expect finite-size scaling to take the form $C(\ell, N) = N^\zeta \Phi(\ell/N)$, where $\Phi$ is a scaling function. In order to estimate $\zeta$, we simulated different systems from $N = 8$, through 256, close to the threshold $F_c$. The chains were released from random initial conditions until they are (barely) pinned. Disorder averages were performed over many independent realizations, from 5000 for smaller $N$'s to 100 for the largest $N$. The insets in Fig. 4 show numerical data for $C(\ell, N)$ as a function of $N$ for fixed $\ell/N$. The slopes of these curves yield the exponent $\zeta$. To find the asymptotic exponent value, we cal-culated the effective exponent at finite $N$ according to $\zeta(N) = \log\left[C(\ell, N)/C(\frac{\ell}{2}, \frac{N}{2})\right]/\log(2)$, and then averaging over all available values of $\ell$ (see Fig. 4). The saturated (large $N$) regime yields $\zeta \approx 1.46$ for uniform chains, and $\zeta \approx 1.22$ for random chains.

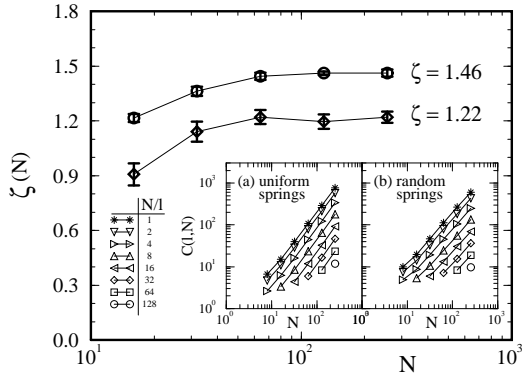

FIG. 4. Effective roughness exponent $\zeta(N)$.

Our value of $\zeta$ for uniform chain is in very good agreement with the FRG result [2] of $\zeta = 3/2$ which is expected to be exact. The result for random chain is also very close to the one reported by Leschhorn and Tang [7] for the driven DP in random-bond disorder ($\zeta \simeq 1.25$).

Our numerical results combined with analytic argument have demonstrated clearly that the critical dynamics of a 1-dimensional random chain in contact with a 1-dimensional random substrate belongs to the universality class of a directed path in 2-dimensional random media. We can thus expect the random chain to possess all of the scaling properties known for the DP, including the power-law distribution of avalanches, which means in this case a wide range of stick-slip events [26]. Our results on a random chain can be readily generalized to type (ii) systems in higher dimensions. Relevance of the latter to a variety of other problems including gel electrophoresis and pattern recognition will be discussed elsewhere. It seems hopeful that the disorder-dominated dynamics and statics of a generic type (ii) system may be understood by making analogy to the appropriate type (i) system.

We benefited from insightful comments from D. S. Fisher. Helpful discussions with M. Kardar and D. R. Nelson are also gratefully acknowledged. This work is supported by ONR-N00014-95-1-1002 through the ONR Young Investigator Program. TH acknowledges additional support by the A. P. Sloan Foundation.